\documentclass[prl,aps,onecolumn,10pt,superscriptaddress]{revtex4-2}
\usepackage{amsmath,amsfonts,amssymb,graphics,graphicx,epsfig,color,times}
\usepackage[utf8x]{inputenc}
\usepackage{color}
\usepackage{bbm, dsfont} 
\usepackage{subfigure}
\usepackage{mathrsfs}
\usepackage{verbatim}
\usepackage{setspace}
\usepackage{esint}
\usepackage{lineno}
\usepackage[unicode=true]
 {hyperref}
\begin{document}
\title{Generating scalable graph states in an atom-nanophotonic interface}
\author{C.-H. Chien}
\affiliation{Department of Physics, National Taiwan University, Taipei 10617, Taiwan}
\affiliation{Institute of Atomic and Molecular Sciences, Academia Sinica, Taipei 10617, Taiwan}
\author{S. Goswami}
\affiliation{Institute of Atomic and Molecular Sciences, Academia Sinica, Taipei 10617, Taiwan}
\author{C.-C. Wu}
\affiliation{Institute of Atomic and Molecular Sciences, Academia Sinica, Taipei 10617, Taiwan}
\author{W.-S. Hiew}
\affiliation{Department of Physics, National Taiwan University, Taipei 10617, Taiwan}
\affiliation{Institute of Atomic and Molecular Sciences, Academia Sinica, Taipei 10617, Taiwan}
\author{Y.-C. Chen}
\affiliation{Institute of Atomic and Molecular Sciences, Academia Sinica, Taipei 10617, Taiwan}
\author{H. H. Jen}
\email{sappyjen@gmail.com}
\affiliation{Institute of Atomic and Molecular Sciences, Academia Sinica, Taipei 10617, Taiwan}
\affiliation{Physics Division, National Center for Theoretical Sciences, Taipei 10617, Taiwan}

\date{\today}
\renewcommand{\r}{\mathbf{r}}
\newcommand{\f}{\mathbf{f}}
\renewcommand{\k}{\mathbf{k}}
\def\p{\mathbf{p}}
\def\q{\mathbf{q}}
\def\bea{\begin{eqnarray}}
\def\eea{\end{eqnarray}}
\def\ba{\begin{array}}
\def\ea{\end{array}}
\def\bdm{\begin{displaymath}}
\def\edm{\end{displaymath}}
\def\red{\color{red}}
\pacs{}
\begin{abstract}
Scalable graph states are essential for measurement-based quantum computation and many entanglement-assisted applications in quantum technologies. Generation of these multipartite entangled states requires a controllable and efficient quantum device with delicate design of generation protocol. Here we propose to prepare high-fidelity and scalable graph states in one and two dimensions, which can be tailored in an atom-nanophotonic cavity via state carving technique. We propose a systematic protocol to carve out unwanted state components, which facilitates scalable graph states generations via adiabatic transport of a definite number of atoms in optical tweezers. An analysis of state fidelity is also presented, and the state preparation probability can be optimized via multiqubit state carvings and sequential single-photon probes. Our results showcase the capability of an atom-nanophotonic interface for creating graph states and pave the way toward novel problem-specific applications using scalable high-dimensional graph states with stationary qubits. 
\end{abstract}
\maketitle

Graph states, as well as cluster states, represent a class of multipartite entangled states \cite{Briegel2001, Hein2004}, which provides the resource and lays the foundation for one-way quantum computation \cite{Raussendorf2001, Walther2005, Briegel2009}. This computing scheme builds upon an irreversible flow of measurement outcomes and can be directly mapped to circuit quantum computation with corresponding single- and two-qubit gate operations \cite{Walther2005}. Recent progresses have demonstrated generation of the cluster states in diverse platforms including stationary and discrete-variable qubits of a superconducting processor \cite{Gong2019, Cao2023}, trapped ions \cite{Lanyon2013}, trapped atoms in an optical tweezer array \cite{Bluvstein2022}, and flying photonic qubits encoded in light polarizations \cite{Walther2005, Kiesel2005, Lu2007, Tokunaga2008, Schwartz2016, Thomas2022, Yang2022} or continuous variables \cite{Larsen2019, Asavanant2019}. For sequentially-woven linear cluster states in time-bin photons \cite{Thomas2022, Yang2022}, only a single setup of single-photon source is required, which is resource efficient but suffers from limited single-photon generation efficiency and state fidelity as the number of qubits scales up. 

One of the solutions to improve the generation efficiency is to harness the strong atom-photon coupling in a cavity \cite{Yang2022}. This strong-coupling quantum interface has recently shown collective enhancement strength \cite{Samutpraphoot2020, Liu2023}, collectively induced transparency \cite{Lei2023}, and improved metrological gain \cite{Li2023}. The photon-mediated interactions within such systems can further allow four-mode square graph states engineering \cite{Cooper2022} or W states preparations \cite{Hiew2023} via state carving technique \cite{Sorensen2003, Chen2015, Welte2017}. In particular, an atom-nanophotonic interface \cite{Chang2018, Corzo2019, Sheremet2023} presents one of the strongly-coupled systems that can host all-to-all dipole-dipole interactions \cite{Solano2017}, chiral spin-exchange couplings \cite{Mitsch2014, Bliokh2014, Pichler2015, Lodahl2017}, intriguing collective radiations \cite{Tudela2013, Mahmoodian2018, Albrecht2019, Jen2020_subradiance, Mahmoodian2020, Jen2021_bound, Jen2022_correlation, Pennetta2022, Pennetta2022_2}, and topological waveguide quantum electrodynamics \cite{Kim2021}. 

Moreover, a versatile and scalable nanophotonic interface with atoms in optical tweezers envisions a transportable entanglement \cite{Dordevic2021}, which promises a coherent atomic quantum processor. Here we propose a scalable generation of arbitrary two-dimensional graph states, including square and linear cluster states, using an atom-nanophotonic cavity platform. We find that they can be prepared with high fidelity by applying the state carving technique. Upon the detection of single photon reflection, the cluster states can be generated subject to single-qubit rotations and can be scaled up to arbitrary sizes under adiabatic transport. We provide a generic protocol to carve out unwanted state components based on the huge contrast of single-photon reflectivity. We also demonstrate the generation of a square graph state in the same setup, which promises a potential application in, for example, the measurement-based quantum eigensolver \cite{Cao2023, Ferguson2021}. Finally we analyze the state fidelity and propose a sequential single-photon probe to further optimize the cluster state preparation.\\
\\
\textbf{Results}\\
\textbf{Atom-nanophotonic cavity.} In an atom-nanophotonic interface with a single-sided cavity, Bell states can be prepared via state carving technique \cite{Sorensen2003, Chen2015, Welte2017} with contrasted single-photon reflectivity spectra and can be transported to distant places \cite{Dordevic2021}, promising a scalable quantum network. The state carving protocol starts from the initialized states $|++\rangle$ with $N=2$ atoms via single-qubit rotations between the atomic coupled state $|1\rangle=|g\rangle$ and the uncoupled state $|0\rangle$ as qubit spaces, where $|+\rangle\equiv(|0\rangle+|1\rangle)/\sqrt{2}$. This leads to a full expansion of four atomic states $(|00\rangle+|01\rangle+|10\rangle+|11\rangle)/2$ with the number of coupled atoms $N_a$ $=$ $0$, $1$, $1$, and $2$, respectively. Through an atom-cavity coupling at the atomic transition $|g\rangle\rightarrow|e\rangle$ and a relatively high single-photon reflectivity $R$ for at least one of the atoms in the coupled states compared to a low $R$ for $|00\rangle$, a weak-field probe projects $|++\rangle$ into $(|01\rangle+|10\rangle+|11\rangle)$ upon a photon detection. This effectively carves out the unwanted component $|00\rangle$ and heralds the bipartite entanglement. The other unwanted component $|11\rangle$ can be removed as well by another state carving, after single-qubit rotations $|0\rangle\leftrightarrow|1\rangle$ on both qubits, upon again detecting a second probe photon. This two-stage state carving protocol effectively generates an entangled state $(|01\rangle+|10\rangle)/\sqrt{2}$ with a probability up to $1/2$ \cite{Sorensen2003} and an optimal state fidelity depending on the maximal and minimal values of $R$ \cite{Dordevic2021}.  

Here we consider an atom-nanophotonic interface in a single-sided cavity as shown in Fig. \ref{fig1}. The Hamiltonians for this system can be written as ($\hbar$ $=$ $1$ and see Supplementary Note $1$) 
\bea
H_{s}=&&\omega_{c}a^{\dagger}a+\omega_a\sum_{\mu=1}^{N_a}\sigma_\mu^{\dagger}\sigma_\mu+\sum_{\mu=1}^{N_a} g_\mu\cos(k_sx_\mu)\left(a^{\dagger}\sigma_\mu+\sigma_\mu^{\dagger}a\right), \label{Hs}
\eea
where the atoms are placed at the antinodes of the cavity fields with a profile $\cos(k_sx_\mu)$ for significant atom-cavity coupling strengths. The system dynamics of the density matrix $\rho$ can be solved from $\dot{\rho}=-i[H_s,\rho]+\gamma \sum_{\mu=1}^{N_a} D[\sigma_\mu]\rho+\kappa D[a]\rho$. The cavity resonance frequency is $\omega_{c}$ with a photon operator $a$, the atomic transition frequency is $\omega_a$ with dipole operators $\sigma_\mu\equiv|g\rangle_\mu\langle e|$ for the ground state $|g\rangle$ and the excited state $|e\rangle$, and an atom-photon coupling constant is $g$. A number of $N_a$ atoms are coupled with the nanophotonic cavity fields via the evanescent waves above the surface of the waveguide \cite{Lodahl2017, Solano2017}, where $k_s$ denotes the wave vector of the guided mode and $D[c]\rho\equiv c \rho c^\dag-\{c^\dag c,\rho\}/2$ with $c\in(a, \sigma_\mu)$. The total decay rate of the atom is $\gamma\equiv 2|dq(\omega)/d\omega|_{\omega=\omega_a}g_{k_s}^2L$, with $|dq(\omega)/d\omega|$ the inverse of group velocity, $q(\omega)$ the resonant wave vector, the coupling strength $g_{k_s}$, and the quantization length $L$. A measure of single-atom cooperativity $C\equiv 4g^2/(\kappa\gamma)$ further identifies a strong-coupling regime $C\gg 1$ \cite{Samutpraphoot2020} under a total cavity decay rate $\kappa=\kappa_{\rm wg}+\kappa_{\rm sc}$ involving the decay rate to the waveguide (wg) and the nonguided rate for scattered (sc) light. This atom-nanophotonic interface can be implemented either with trapped atoms in optical lattices \cite{Corzo2019} or in optical tweezer arrays \cite{Samutpraphoot2020, Dordevic2021} as shown schematically in Fig. \ref{fig1}.  
\begin{figure}[!t]
\centering{}\includegraphics[width=0.45\textwidth]{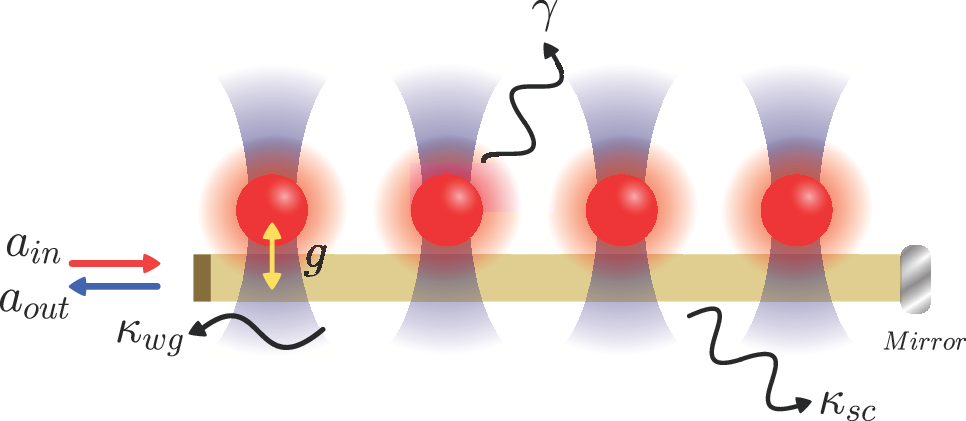}
\caption{\label{fig1}\textbf{Schematic plot of an atom-nanophotonic cavity.} An optical tweezer array of atoms trapped near the waveguide presents a strongly-coupled quantum interface. A single-photon reflectivity measurement can be conducted via collecting the output photon $a_{out}$ from an input one $a_{in}$ through the atom-photon interactions in a single-sided cavity. An atom-cavity coupling constant is $g$ with a cavity decay rate to the waveguide $\kappa_{wg}$ and a loss rate for scattered photons $\kappa_{sc}$. The $\gamma$ quantifies the atomic spontaneous emission rate.}
\end{figure}

\begin{figure}[!b]
\centering{}\includegraphics[width=0.7\textwidth]{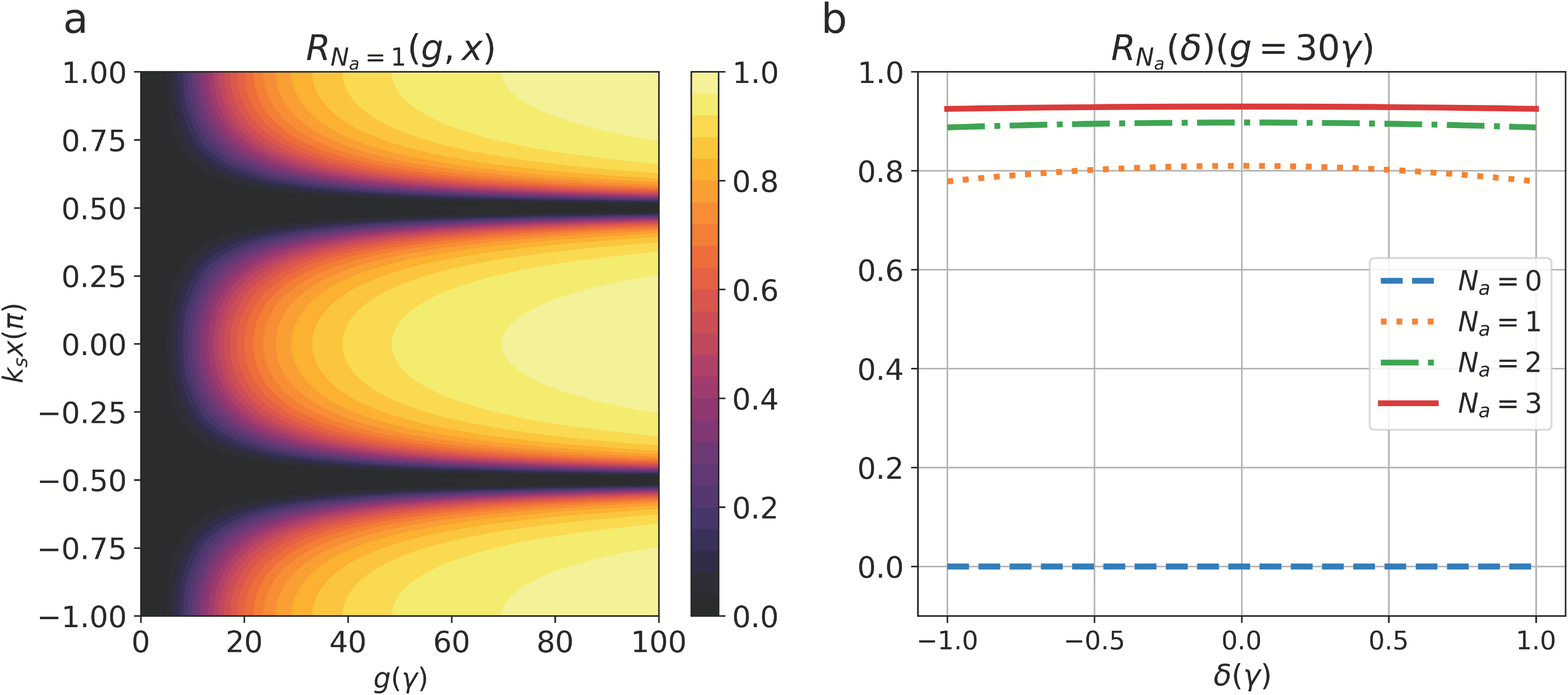}
\caption{\label{fig2}\textbf{Single-photon reflectivity.} (a) The single-photon reflectivity $R$ for $N_a=1$ coupled atom is plotted for various $k_sx_\mu$ in the cavity field profile $\cos(k_sx_\mu)$ and for various atom-cavity coupling $g$, under a resonant laser $\delta\equiv\omega-\omega_c=0$ at the critical coupling regime $\kappa_{\rm wg}=\kappa_{\rm sc}=200\gamma$. As $g/\gamma\geq 20$, the on-resonance $R$ approaches its maximum and saturates as $g$ increases, where it becomes more immune to the position variations from the antinodes of the cavity field profile. (b) In this critical coupling regime, a vanishing $R$ for $N_a=0$ emerges and allows the contrasted reflectivity spectrum compared to the ones for a finite coupled number of atoms $N_a$ under the parameters $(g, \kappa_{\rm wg}, \kappa_{\rm sc})/\gamma = (20, 200, 200)$ and $k_sx_\mu = \pm\pi$.}
\end{figure}

\begin{figure}[!b]
\centering{}\includegraphics[width=0.7\textwidth]{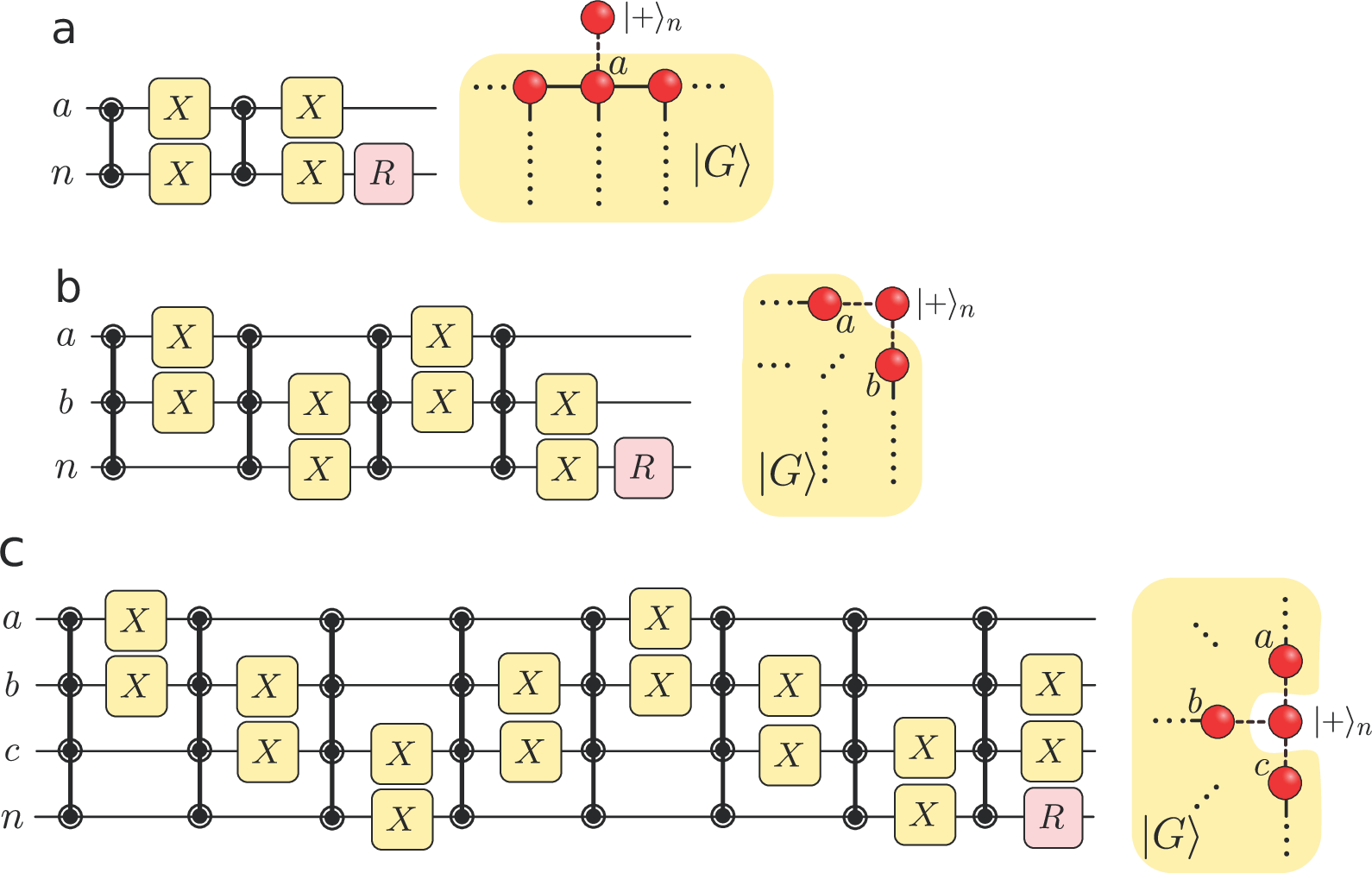}
\caption{\label{fig3}\textbf{Weaving graph states in one and two dimensions.} The graph states can be generated with state carving on the extra qubit $|+\rangle_n$ and (a) the qubit node $a$, (b) the qubit nodes $a$ and $b$, and (c) the nodes of $a$, $b$, and $c$, in the graph state $|{\rm G}\rangle$. These represent multiqubit state carvings on $N_a=2$, $3$, and $4$, respectively. The state carving protocol in (a) provides the foundation to generate the linear cluster states, while all protocols in (a,b,c) and the multiqubit state carving on a larger $N_a$ can generate two-dimensional graph states at arbitrary sizes and shapes. The partially-filled circles with solid links denote the state-carving protocol for $N_a$ atoms. $X$ and $R$ represent the Pauli-$X$ gate and a rotation operation with an angle of $\pi/2$ along the $y$ axis, respectively.}
\end{figure}

The single-photon reflectivity $R=|r|^2$$=$$|a_{\rm out}/a_{\rm in}|^2$ can be obtained by solving the steady-state solutions of $\langle A\rangle={\rm Tr}[\rho(t\rightarrow\infty) A]$ where $A\in\{a,\sigma_1,\sigma_2,...,\sigma_{N_a}\}$ based on the input-output formalism \cite{Caneva12015} under a weak excitation limit ($|g\rangle_\mu\langle g|\approx 1$) with $a_{\rm out}+a_{\rm in}=\sqrt{\kappa_{\rm wg}}a$ (see Methods and Supplementary Note $1$). In Fig. \ref{fig2}, we present the on-resonance single-photon reflectivity $R$ for two coupled atoms $N=2$ and the reflectivity spectrum for different $N$. We find that when a critical coupling regime is reached \cite{Samutpraphoot2020}, that is $\kappa_{\rm wg}=\kappa_{\rm sc}$ (Supplementary Note $2$), a vanishing $R$ emerges (Figs. \ref{fig2}a and \ref{fig2}b). This suggests the most contrasted single-photon reflectivity which can be utilized to conduct the state carving and project the target states with high fidelity. Meanwhile, this critical coupling regime can be manipulated and controlled by the number of air holes in the region of mirror in the photonic crystal cavity \cite{Chan2009, Groblacher2013, Samutpraphoot2020}. We note that the on-resonance $R$ in Fig. \ref{fig2}a are robust to the position variations from the anti-nodes of the cavity field profile in the strong coupling regime. On the other hand in the moderate coupling regime, a sensitivity to position fluctuations around the anti-nodes $k_sx_\mu=\pm\pi$ arises \cite{Samutpraphoot2020}, which leads to a degradation in the generation probability of graph states but not the fidelity. This fragility can be mitigated by lower the temperature of the atoms in optical tweezers via Raman sideband cooling \cite{Kaufman2012} or gray-molasses loading to optical tweezers \cite{Brown2019}. While an adiabatic transport of atoms can manipulate the coupling or uncoupling between the atoms and the cavity photon, which may suffer from atom losses or decoherences \cite{Dordevic2021}, an extra qubit state $|2\rangle$ uncoupled to the cavity can be utilized and transferred from $|1\rangle$ to decouple from the cavity. This only requires a single-qubit gate between $|1\rangle$ and $|2\rangle$, and should be more resilient to adiabatic transfer errors.\\
\\
\textbf{Generation of linear and two-dimensional graph states.} With the contrasted single-photon reflectivity spectrum in the atom-nanophotonic cavity interface, here we propose to generate a special class of multipartite entangled states including linear and two-dimensional cluster states (Fig. \ref{fig3}). These entangled states establish the foundations for measurement-based quantum computation \cite{Raussendorf2001, Walther2005, Briegel2009, Raussendorf2003, Larsen2021}. We express the target cluster states in the conventional graph representations, where the nodes or vertices are linked by edges, demonstrating the individual qubits $|+\rangle$ and the controlled-Z (CZ) gates, respectively. For two-node and three-node linear cluster states as an example, they are $(|+0\rangle+|-1\rangle)/\sqrt{2}$ and $(|+0+\rangle+|-1-\rangle)/\sqrt{2}$, respectively, as $|\pm\rangle\equiv(|0\rangle\pm|1\rangle)/\sqrt{2}$, and CZ gate imprints an extra $\pi$ phase conditional on the two-qubit states $|11\rangle$, transforming $|1\pm\rangle$ to $|1\mp\rangle$. For four-node linear cluster state, it can be expressed as $|\psi\rangle_{LCS}$ $=$ $(|+0+0\rangle+|+0-1\rangle+|-1-0\rangle+|-1+1\rangle)/2$, where a notation of $|({\rm sign})({\rm digit})({\rm sign})({\rm digit})...\rangle$ is adopted. Before applying the state carving (SC) protocol for preparing scalable linear graph states $|{\rm LGS}\rangle$, it would be convenient to rotate them to what we dub the `precursor' state $(|1010\rangle+|1001\rangle+|0100\rangle+|0111\rangle)/2$ by single-qubit rotations $R$ on the qubit number $1$ and $3$, respectively (Supplementary Note $3$).  

The conventional linear cluster state generation requires a CZ gate on two qubits linking one of the nodes denoted by $a$ in the initial graph state $|{\rm G}\rangle$ with an extra qubit $|+\rangle_n$. The resultant new graph state becomes 
\bea
{\rm CZ}(a,n)|{\rm G}\rangle|+\rangle_n=\frac{1}{\sqrt{2}}\left(|\phi_0\rangle|0\rangle_a|+\rangle_n+|\phi_1\rangle|1\rangle_a|-\rangle_n\right),\label{c}
\eea
where ${\rm CZ}(a,n)$ represents the CZ gate operation and the states $|\phi_{0(1)}\rangle$ are the graph state components associated with the qubit node $a$. The resultant graph state is the target state for our proposed protocols using SC in an atom-nanophotonic interface. With a vanishing $R$ for the states $|00\rangle$, we obtain the projected and normalized states after SC$(a,n)$ between the qubits $a$ and $n$ upon single-photon reflection, 
\bea
{\rm SC}(a,n)|{\rm G}\rangle|+\rangle_n&&={\rm SC}(a,n)\left(|\phi_0\rangle|0\rangle_a|0\rangle_n+|\phi_0\rangle|0\rangle_a|1\rangle_n+|\phi_1\rangle|1\rangle_a|0\rangle_n+|\phi_1\rangle|1\rangle_a|1\rangle_n\right),\nonumber\\
&&\rightarrow \frac{1}{\sqrt{3}}\left(|\phi_0\rangle|0\rangle_a|1\rangle_n+|\phi_1\rangle|1\rangle_a|0\rangle_n+|\phi_1\rangle|1\rangle_a|1\rangle_n\right),\label{sc} 
\eea  
where $|1\rangle_a|1\rangle_n$ can be rotated to $|0\rangle_a|0\rangle_n$ by single-qubit operations and carved out again using SC, transforming Eq. (\ref{sc}) to the target state of Eq. (\ref{c}) up to single-qubit rotations as shown in Fig. \ref{fig3}a (Supplementary Note $4$). This lays the foundation of the protocol for any linear cluster states by weaving the cluster state with extra qubits successively [$2(M-1)$ times of SC for $M$ nodes] and for certain two-dimensional graph states like a cross with five nodes or horseshoe cluster states \cite{Walther2005}.

To weave general two-dimensional cluster states, we need multiqubit SC on qubits more than $N_a=2$ (Figs. \ref{fig3}b and \ref{fig3}c). We take $N_a=3$ in Fig. \ref{fig3}b as an example, and the target state to link the extra qubit $|+\rangle_n$ with two edges and two nodes ($a$ and $b$) in the graph state becomes
\bea
{\rm CZ}(a,n){\rm CZ}(b,n)|{\rm G}\rangle|+\rangle_n=\frac{1}{2}\left(|\phi_{00}\rangle|00\rangle_a|+\rangle_n + |\phi_{01}\rangle|01\rangle_a|-\rangle_n + |\phi_{10}\rangle|10\rangle_a|-\rangle_n + |\phi_{11}\rangle|11\rangle_a|+\rangle_n\right).
\eea
Via four times of SC on the nodes $a$, $b$, and $n$ to remove the state component $\rangle|00\rangle_a|0\rangle_n$, followed by $X$ gates on ($a$, $b$) and ($b$, $n$) respectively and successively as in Fig. \ref{fig3}b, we obtain (Supplementary Note $4$)
\bea
\frac{1}{2}\left(|\phi_{00}\rangle|00\rangle_a|1\rangle_n + |\phi_{01}\rangle|01\rangle_a|0\rangle_n + |\phi_{10}\rangle|10\rangle_a|0\rangle_n + |\phi_{11}\rangle|11\rangle_a|1\rangle_n\right),
\eea
which is exactly the target precursor state and then can be transformed to the target graph state by applying a single-qubit rotation on the $n$th qubit, fulfilling the protocol to generate graph states more than one dimension (Figs. \ref{fig3}b and \ref{fig3}c). In general, an arbitrary $M$ links with $|+\rangle_n$ can be achieved by applying the similar designs in the protocols of Fig. \ref{fig3}. Two arbitrary graph states, $|{\rm G}_1\rangle$ and $|{\rm G}_2\rangle$, can also be combined by SC on an additional qubit $|+\rangle_n$ and two nodes in respective graph states. This can be extended to combine arbitrary number of $M$ graph states into a larger graph state with again SC on an additional $|+\rangle_n$ and $M$ nodes with $2^M$ numbers of SC (Supplementary Note $4$).

\begin{figure}[!b]
\centering{}\includegraphics[width=0.65\textwidth]{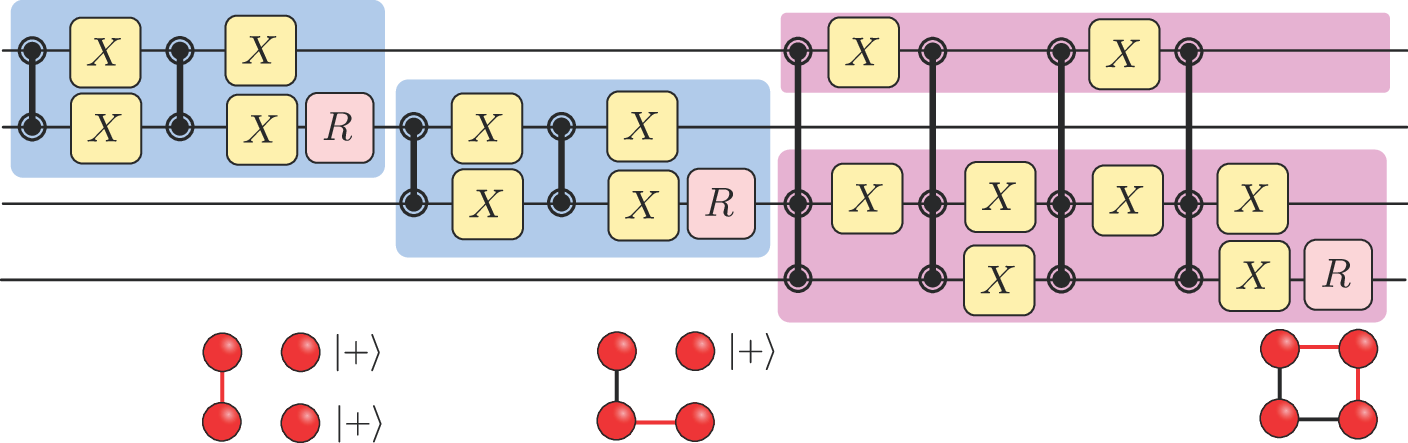}
\caption{\label{fig4}\textbf{Square graph state preparation.} A series of state carvings on one-link graph state (blue shades) and two-link graph state (purple shades) weave the initialized states $|+\rangle^{\otimes 4}$ into the square graph state, where intermediate stages of the generated graph states are shown below the circuits.}
\end{figure}

\begin{figure}[!t]
\centering{}\includegraphics[width=0.99\textwidth]{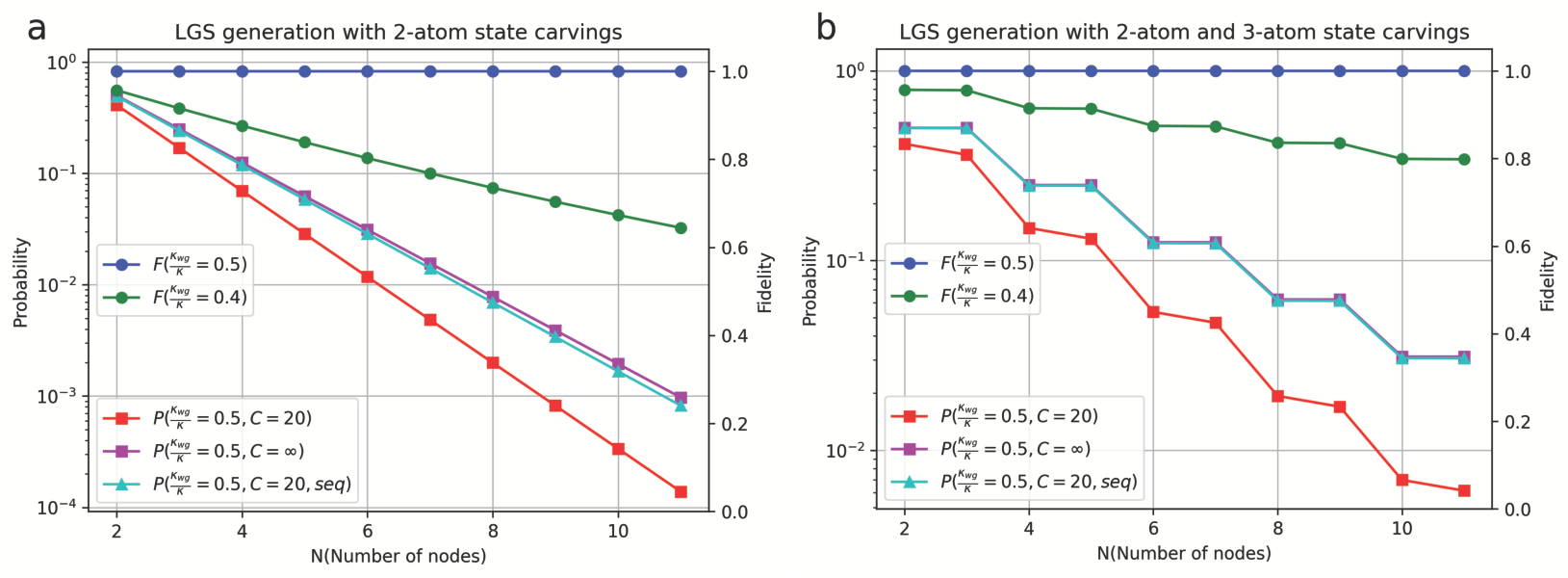}
\caption{\label{fig5}\textbf{Probability and fidelity of linear graph states (LGS) generations.} Two different protocols using $2$-atom and multi-atom state carvings are shown in (a) and (b), respectively. The probability at a finite $C$ decreases as the number of nodes $N$ increases, which can be improved for two sequential single-photon probe ($N_p=2$), approaching the optimized result at an infinite $C$. The corresponding fidelity is perfect under the critical coupling regime at $\kappa_{wg} = \kappa_{sc}  = \kappa/2$, which degrades as deviating from this regime.}
\end{figure}

One of the basic elements in two-dimensional graph states is the square graph state. This can be weaved from two one-link qubits and one two-link qubit (Fig. \ref{fig4}), which results in $(|0+0+\rangle+|0-1-\rangle+|1-0-\rangle+|1+1+\rangle)/2$. A generic two-dimensional graph state at arbitrary sizes and shapes can then be realized with assistance of multiqubit SC as demonstrated in Fig. \ref{fig3}. On the other hand, linear graph states can be created only by using $2$-atom SC [Fig. \ref{fig3}(a)], but more exquisite protocols can as well be developed involving multiqubit SC that can improve the optimal probability to generate them, which we discuss more in the next section and in Supplementary Note $4$.\\ 
\\
\textbf{Probability and fidelity of graph state preparation.} In the following, we investigate the probability and the fidelity of graph state preparation, which can provide insights into the capabilities and limits of the proposed protocols. We also propose an improving protocol by applying a sequential single-photon probe to create optimal probability, high-fidelity, and scalable graph states. A trade-off for our protocol in Fig. \ref{fig3}(a) is the reduction of the generation probability with the size of the graph state, as more unwanted states need to be carved away. In this regard, we utilize either the multiqubit SC between one node and $|+\rangle^{\otimes 2}$ or multiqubit SC among $|+\rangle^{\otimes N}$ specifically with the digit-sign representation (alternative to sign-digit representation) in the target linear graph states, which optimizes the probability of state preparations with less state components to be carved out (Supplementary Note $4$). We compare the probability and fidelity for linear graph state generation by $2$-atom (one node and $|+\rangle$) and $3$-atom state carving (one node and $|+\rangle^{\otimes 2}$) protocols in Fig. \ref{fig5}. The maximum probability $P$ of generating linear graph states in each protocol can be achieved for the ideal case of an infinite cooperativity (i.e. $C=\infty$ and $R_{N_a\geq 1}=1$), which yields $P=2^{-(N-1)}$ and $2^{-\lfloor N/2\rfloor}$ with a floor function of $N/2$, respectively, showing the advantage of multiqubit SC. In practice, a finite $C$ would produce a non-unity reflectivity $R$. For a number of $N_{SC}$ state carvings, a factor of $R^{N_{SC}}$ will be multiplied to the probability which quickly decreases as $N$ increases (red squares in Fig. \ref{fig5} for $C = 20$ as an example).  

To achieve the maximum probability, we would need to modify the state carving procedure by sending multiple single photons sequentially for each state carving instead of using just one photon each time. This procedure will dramatically decrease the deteriorating effect of non-unity $R$. For example, the total probability of getting a photon in reflection for Bell state-carving will rise from $R_1/2$ to $[1-(1-R_1)^{N_p}]/2$ for $N_p$ sequential photons (Supplementary Note $5$). If $R_1= 0.8$, with just $N_p =2$ we have $1-(1-R_1)^{N_p} = 0.96$ which is near unity, and this dramatically improves the probability to nearly reach the optimal probabilities at an infinite $C$. 

The fidelity of the generated graph states is a deciding factor for measurement-based quantum applications. A high-fidelity graph state indicates genuine multipartite entanglement, which is essential as a resource for one-way quantum computation. The fidelity for generating arbitrarily large graph states becomes perfect when $\kappa_{wg} = \kappa_{sc}  = \kappa/2$ as the critical coupling regime with $r_0 = 0$ (blue circles in Fig. \ref{fig5}). We note that $r_0$ is the probability of reflection when no atom is coupled to the cavity. All state carvings intend to remove these uncoupled states (i.e., $|00\rangle$, $|000\rangle$, or $|0000\rangle$ states etc.). Hence, if $r_0 =0$, no unwanted state component is projected out when a photon is reflected, resulting in a unit fidelity of the desired state. If $r_0 \neq 0$, some unwanted state contribution emerges, and the fidelity deteriorates as seen in the case of $\kappa_{wg}/\kappa = 0.4$ (green circles in Fig. \ref{fig5}). However, these fidelities can also be improved arbitrarily close to $1$ when multiple sequential photons are used. Furthermore, the fidelity is resilient to uneven atom-photon couplings (i.e., $g_i \neq g_j$ or equivalently $C_i \neq C_j$ for i$^{th}$ and j$^{th}$ atoms). Therefore, our proposed protocol is robust against multiple forms of imperfections of $g_i \neq g_j$, $r_0 \neq 0$ away from the critical coupling regime, and $|r_1| < 1$ for a finite $C$ (Supplementary Note $5$).\\
\\
\textbf{Discussion and conclusion.} Our proposed protocol using an atom-nanophotonic interface provides a high-fidelity generation of scalable graph states. We present a general recipe to weave graph states in one and two dimensions, where we provide a multiqubit state carving for linear and two-dimensional graph states at arbitrary sizes. This exquisite design protocol relies on the feature of contrasted single-photon reflection spectra allowed by the critical coupling regime in the interface. Via the state-carving technique, we are able to project the system into the target graph states with high fidelity. A sequence of single-photon probes further enhances the graph state probability, which is especially useful for large-size graph states and promises a near-term application in quantum engineering of multipartite entangled states. Our results illustrate the potential of an atom-nanophotonic cavity for generating linear and high-dimensional graph states, which sets the foundation for measurement-based quantum computation and many intriguing problem-specific applications.\\
\\
\textbf{Methods}\\
\textbf{\small Atom-photon coupled equations and input-output formalism.}\\
The atom-photon coupled equations can be obtained from the Hamiltonian in Eq. (\ref{Hs}) along with corresponding Lindblad forms. In the Heisenberg picture with the input-output formalism, we have the time-evolving equations of motions, 
\bea
\dot{a}&&=\left(i\delta-\frac{\kappa}{2}\right)a -i\sum_{\mu=1}^{N_a}g_\mu\cos(k_s x_\mu)\sigma_\mu +\sqrt{\kappa_{\rm wg}}a_{\rm in},\\
\dot{\sigma}_\mu&&= \left(i\delta-\frac{\gamma}{2}\right)\sigma_\mu+ig_\mu\cos(k_s x_\mu)a(|e\rangle_\mu\langle e|-|g\rangle_\mu\langle g|), 
\eea
where a photon detuning $\delta\equiv \omega-\omega_c=\omega-\omega_a$ and a Kronecker delta function $\delta_{\mu,\nu}$. With a weak probe field at a frequency $\omega$ and under the weak excitation limit $|g\rangle_\mu\langle g|\approx 1$, we are able to calculate the single-photon reflectivity $R=|a_{\rm out}/a_{\rm in}|^2$ in the steady-state solutions.\\
\\
\textbf{Acknowledgments}\\
We acknowledge support from the National Science and Technology Council (NSTC), Taiwan, under the Grants No. 112-2112-M-001-079-MY3 and No. NSTC-112-2119-M-001-007. We are also grateful for support from TG 1.2 of NCTS at NTU. \\
\\
\textbf{Author contributions}\\
H.H.J. conceived the original idea. C.-H.C. conducted the theoretical analysis and designed the preparation protocol for scalable graph states. S.G. proposed the ideas for calculation of the state fidelity and sequential single-photon probes. Y.-C.C. and H.H.J. supervised the project. All authors participated in discussions and contributed to the writing of the manuscript.

\end{document}